\documentclass[12pt]{article}
\usepackage{graphicx,amssymb}
\usepackage{citesort}

\newcommand{\figmsComp}
{
\begin{figure}[h]
 \centerline{
 \resizebox*{0.58\textwidth}{!}{
 \includegraphics{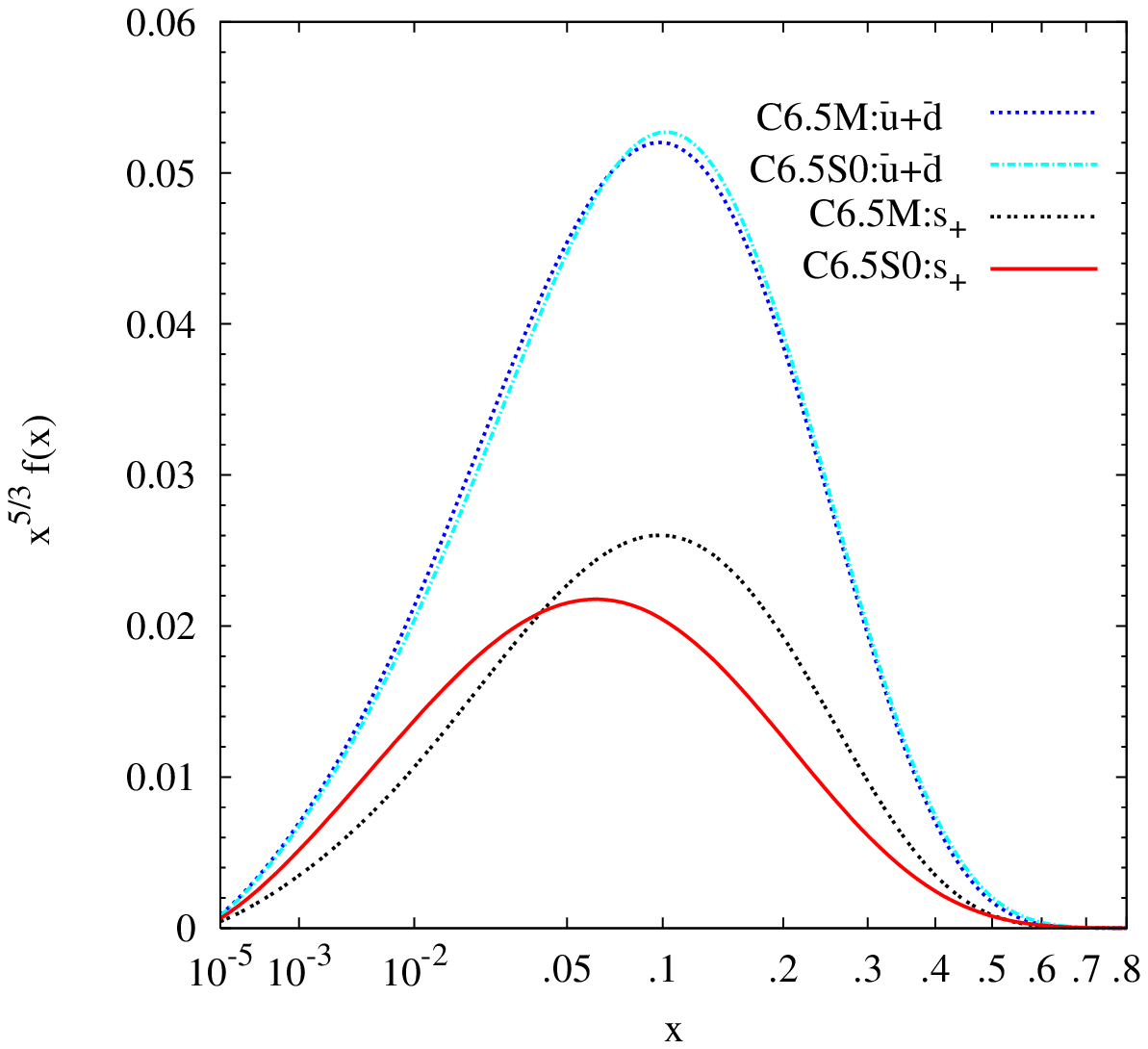}
 }
 }
\caption{The new CTEQ6.5S0 $s_+(x)$ and $\bar{u}(x)+\bar{d}(x)$ PDFs at
$\mu=1.3\ \mathrm{GeV}$ compared to those of CTEQ6.5M.
}%
\label{fig:6m6sComp}
\end{figure}
}
\newcommand{\figChi}
{
\begin{figure}[h]
 \centerline{
 \resizebox*{0.6\textwidth}{!}{
 \includegraphics{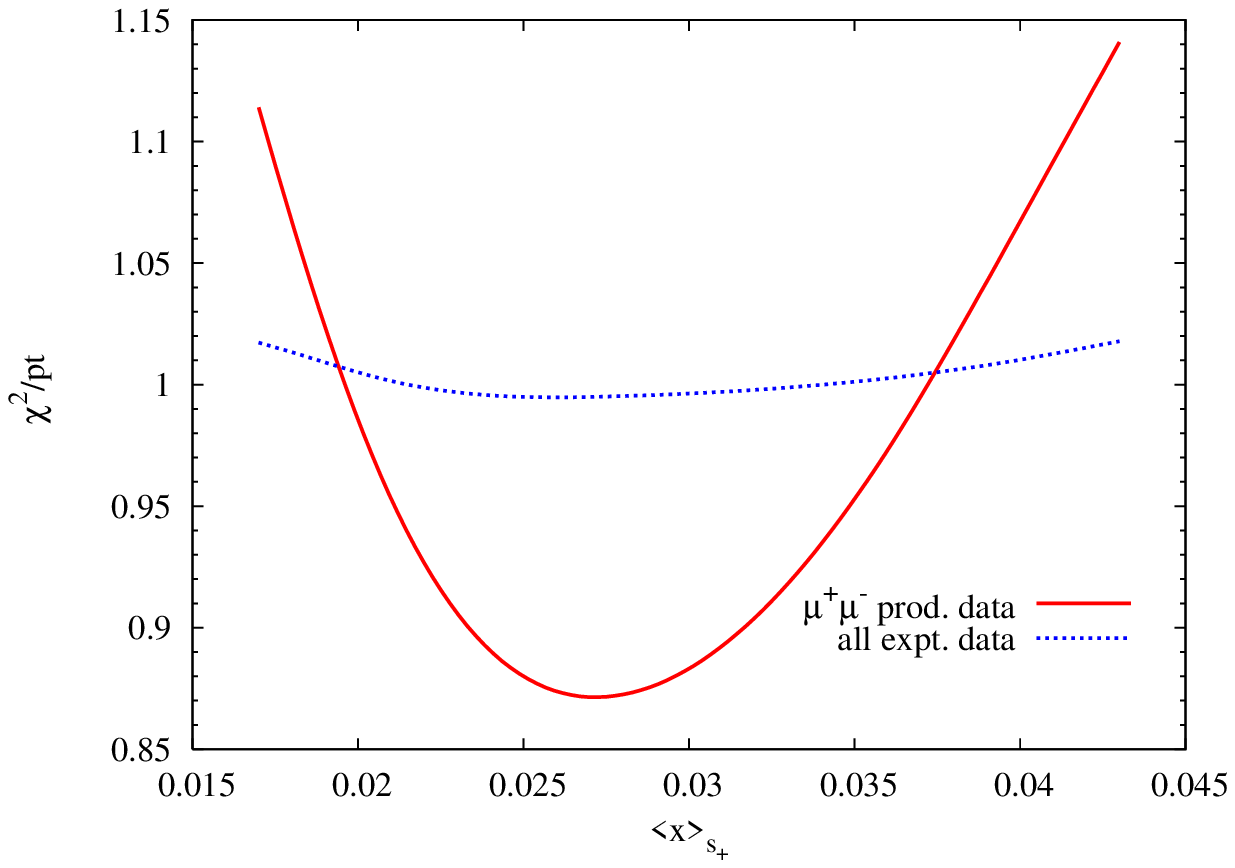}
 }
 }
\caption{$\chi^{2}$ per data point vs. $\langle x\rangle _{s_{+}}$ for the full data set
(3542 points) and for the $\nu$ and $\bar{\nu}$ dimuon production data (149 points).
}%
\label{fig:StrChiKappa}
\end{figure}
}
\newcommand{\figsplus}
{
\begin{figure}[h]
 \centerline{
 \resizebox*{0.6\textwidth}{!}{
 \includegraphics{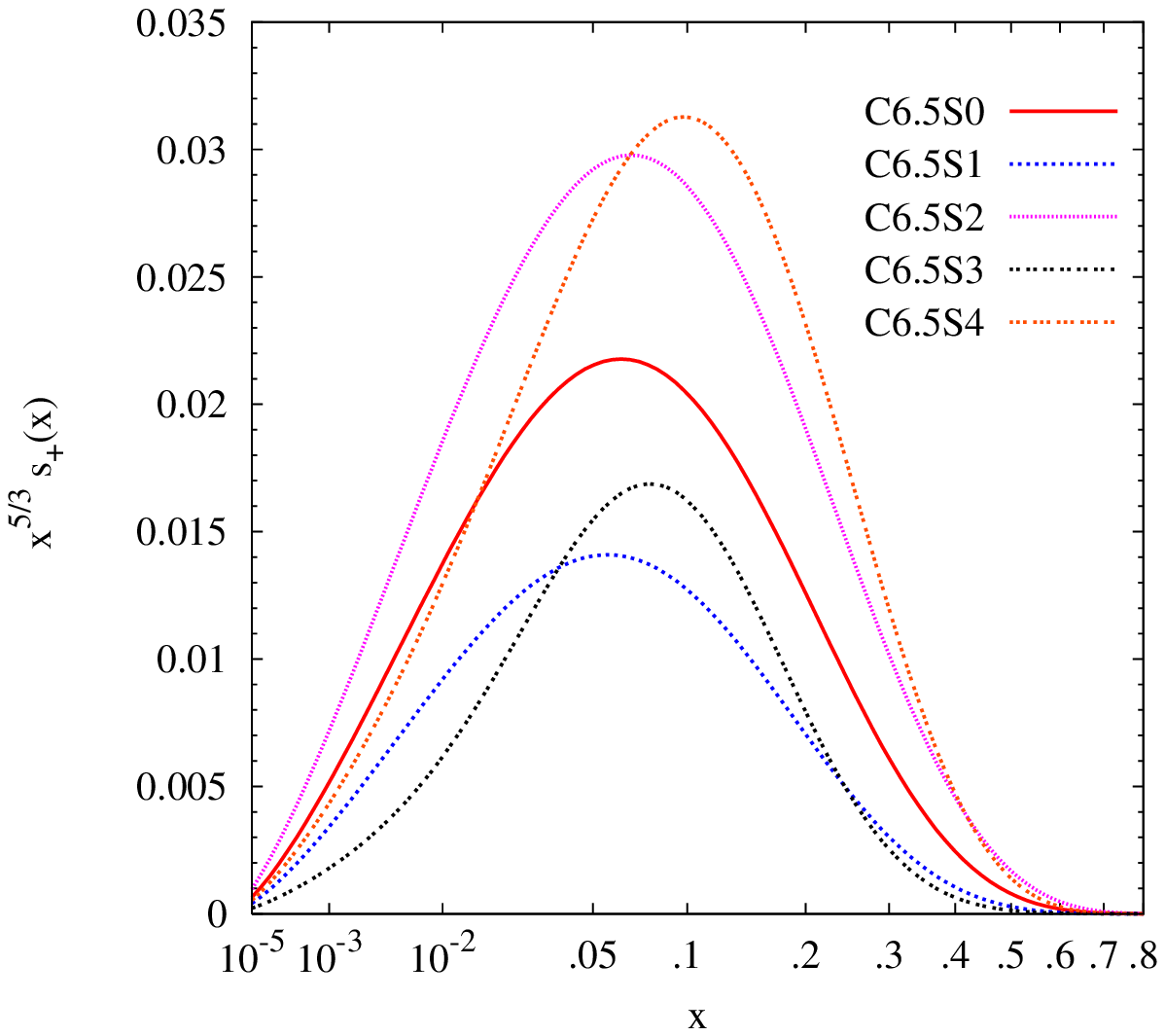}
 }
 }
\caption{The strangeness distribution $s_{+}(x)$ at the initial scale
$\mu =1.3\,\mathrm{GeV}$ for the five PDF sets CTEQ6.5Si, $i=0,\dots,4$.
}%
\label{fig:splus5}
\end{figure}
}
\newcommand{\fignutevS}
{
\begin{figure}[h]
 \resizebox*{0.48\textwidth}{!}{
 \includegraphics{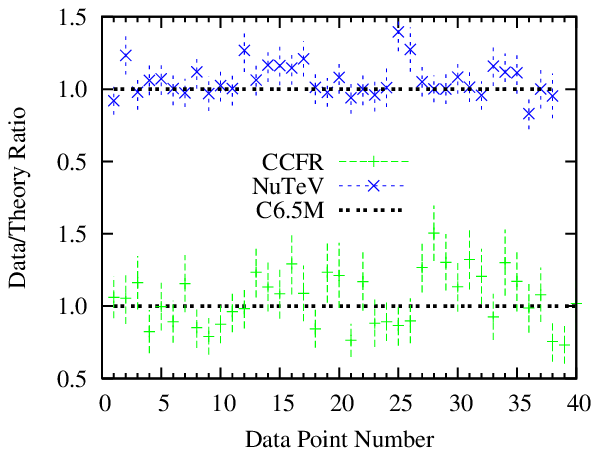}
 } \hfill
 \resizebox*{0.48\textwidth}{!}{
 \includegraphics{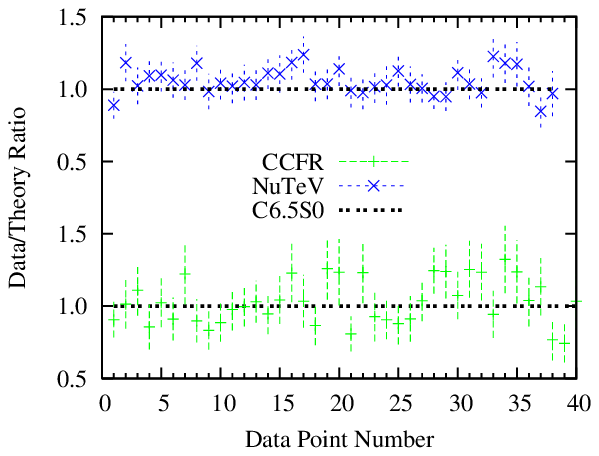}
 }
\caption{Comparison of neutrino dimuon data to fits using CTEQ6.5M (left panel)
 and the new set CTEQ6.5S0 of Sec.\ \ref{sec:c65s0} (right panel). The data points are sorted in $y$-bins,
and within each $y$-bin, by $x$ value, and then $Q$ value.
}%
\label{fig:nutevS}
\end{figure}
}
\newcommand{\figSminus}
{
\begin{figure}[h]
 \resizebox*{0.46\textwidth}{!}{
 \includegraphics{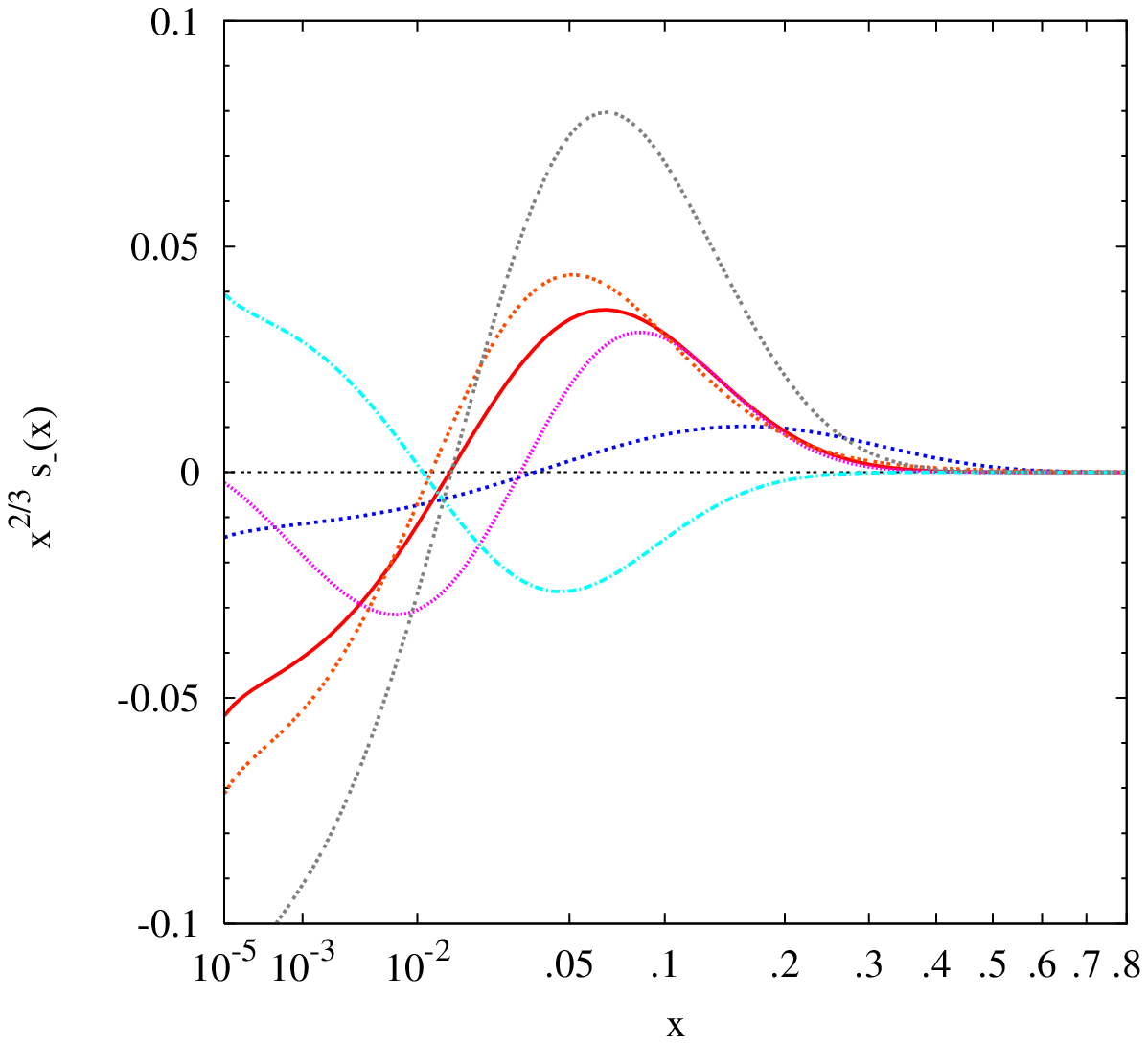}
 } \hfill
 \resizebox*{0.48\textwidth}{!}{
 \includegraphics{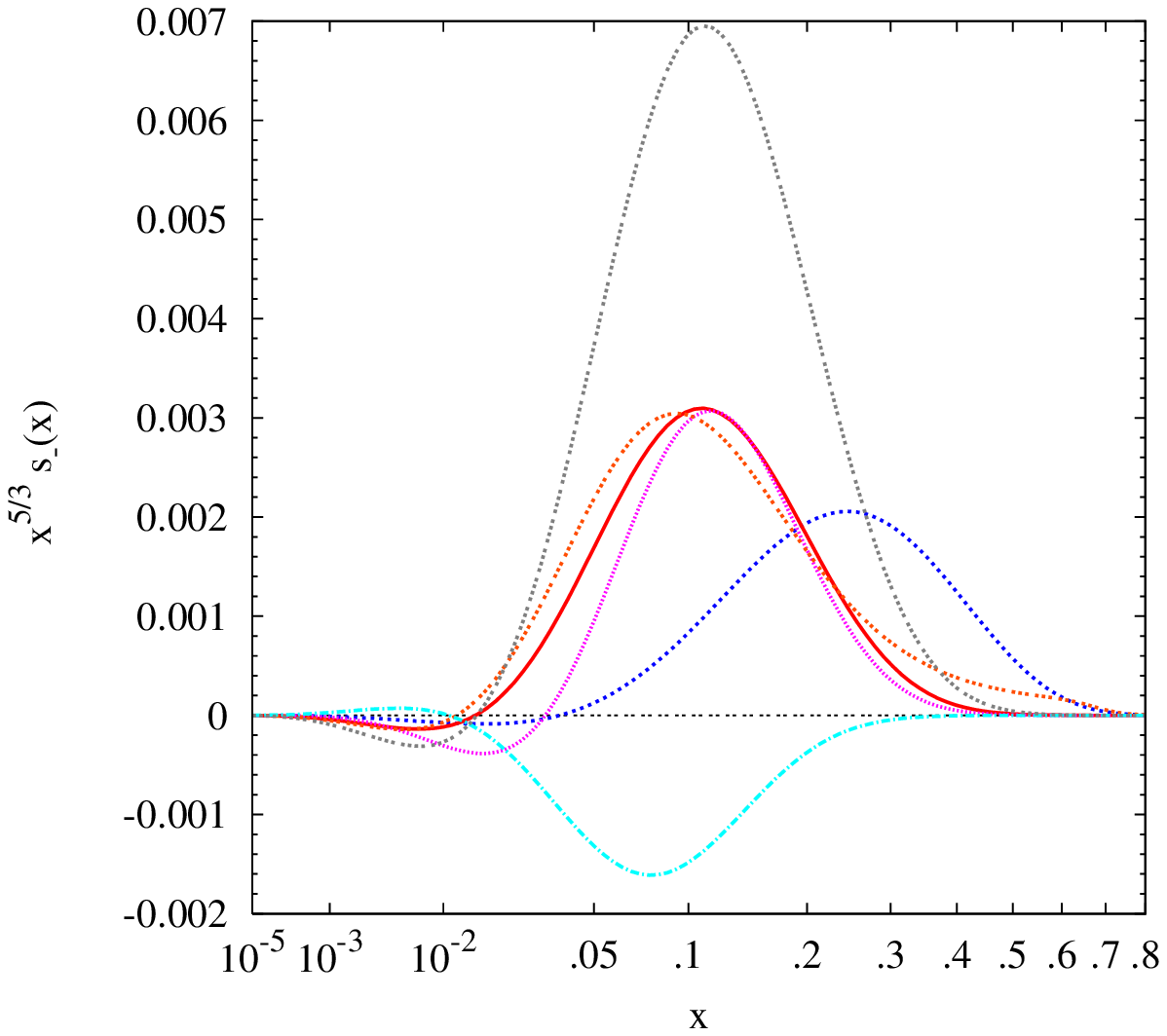}
 }
\caption{Examples of strangeness asymmetry function $s_-(x)$ that
are consistent with existing experimental data (left panel); and the
corresponding momentum distribution $x\,s_-(x)$ (right panel).
}%
\label{fig:sminus}
\end{figure}
}
\newcommand{\figGsWc}
{
\begin{figure}[h]
 \resizebox*{0.49\textwidth}{!}{
 \includegraphics{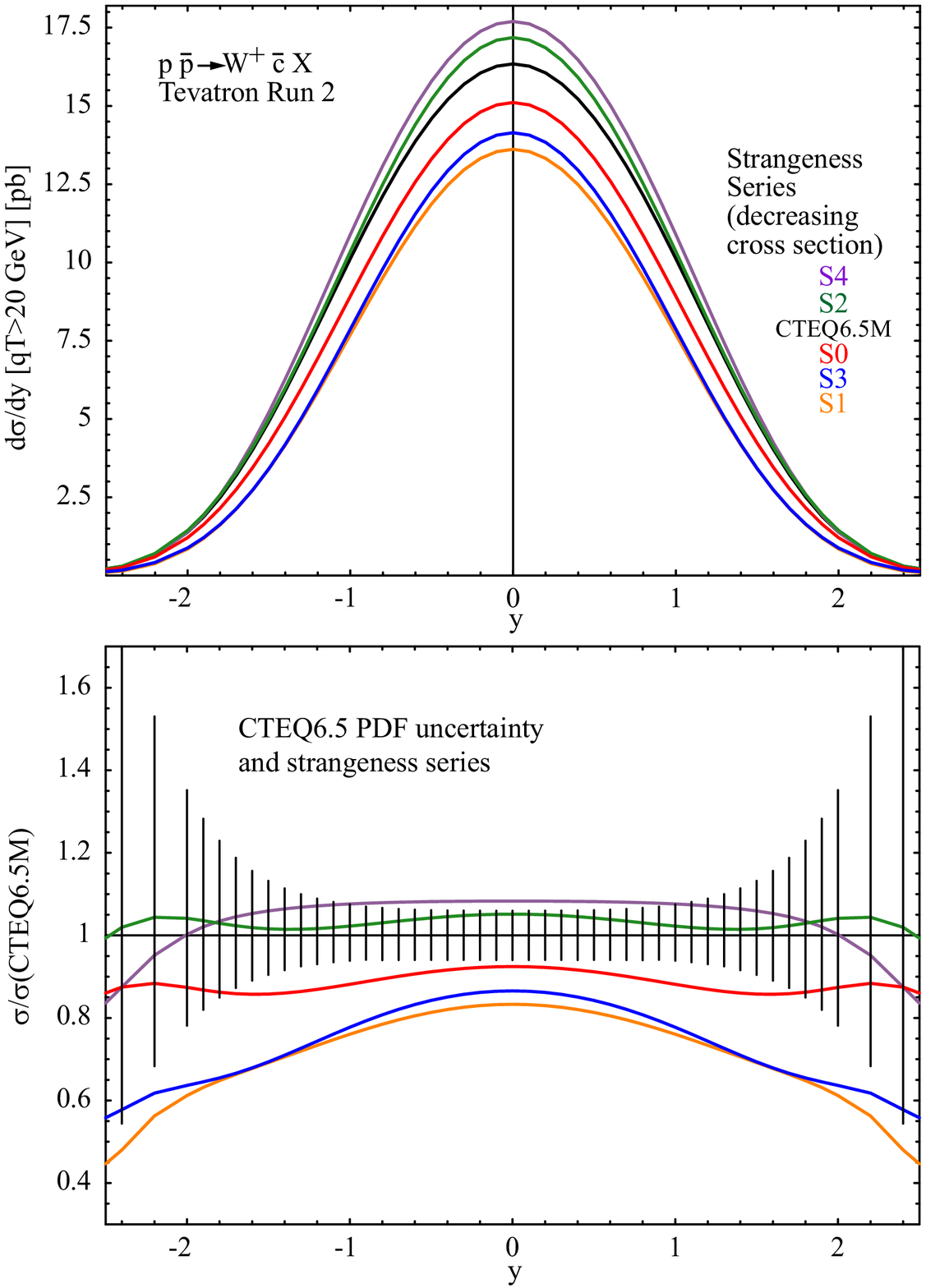}
 } \hfill
 \resizebox*{0.49\textwidth}{!}{
 \includegraphics{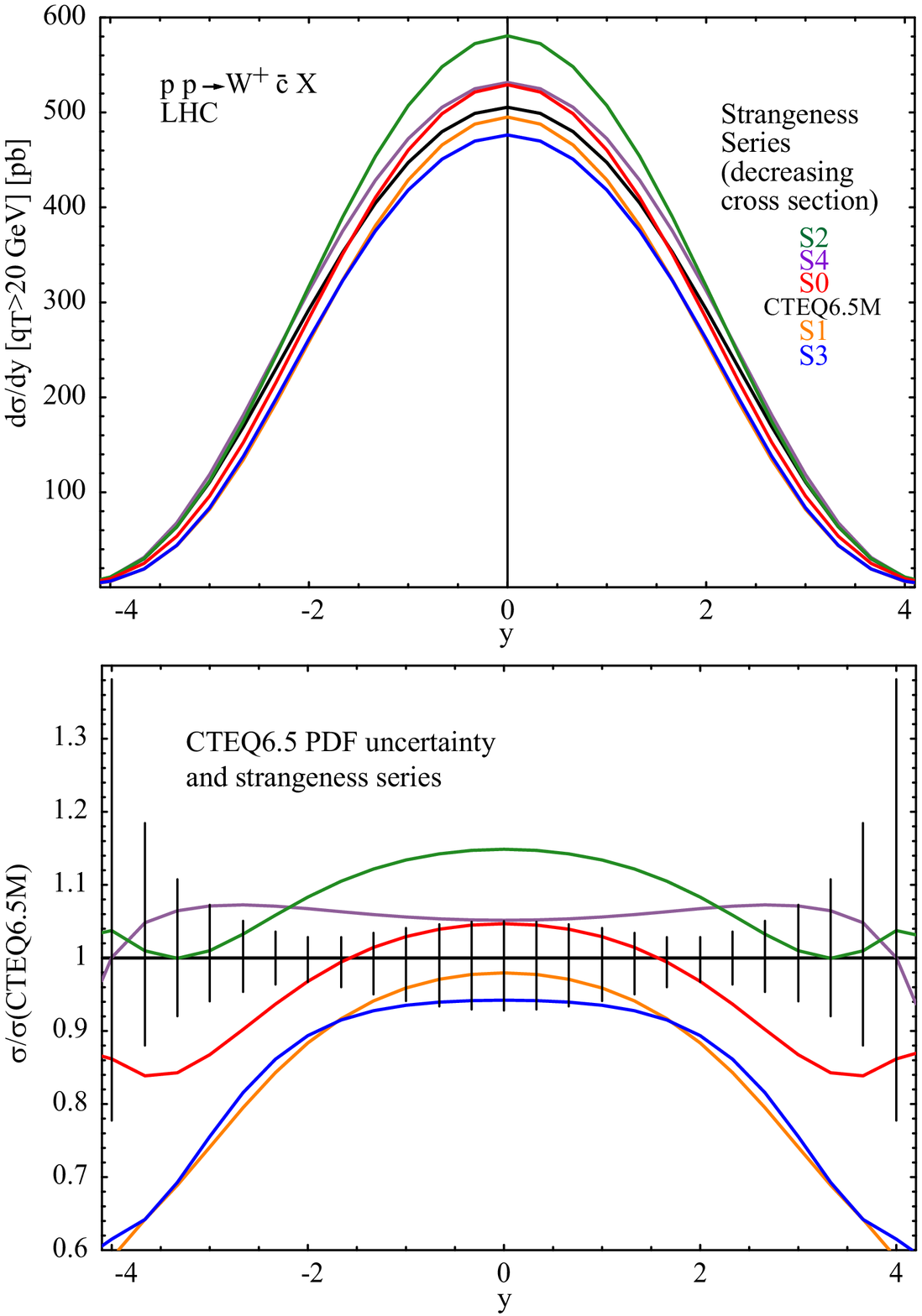}
 }
\centerline{\hspace{2em}(a)\hspace{0.5\textwidth}(b)}
\caption{The rapidity distribution $d\sigma/dy$ at $q_T > 20$ GeV (top
 panel) and its fractional uncertainty (bottom panel) at the Tevatron
 Run-2 (left) and the LHC (right).
}%
\label{fig:gs2wc}
\end{figure}
}
\newcommand{\figHiggs}
{
\begin{figure}[h]
 \resizebox*{0.49\textwidth}{!}{
 \includegraphics{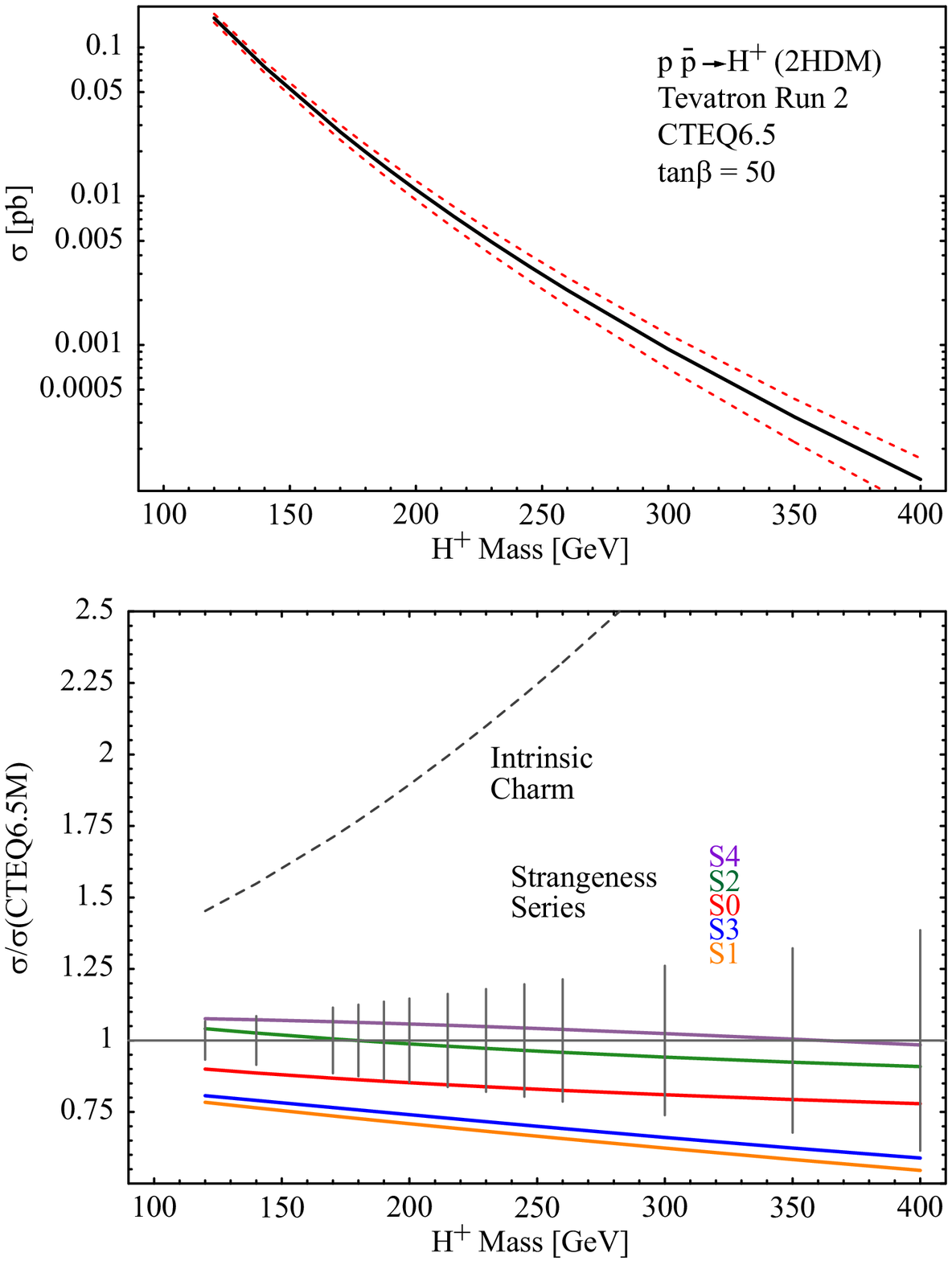}
 } \hfill
  \resizebox*{0.49\textwidth}{!}{
 \includegraphics{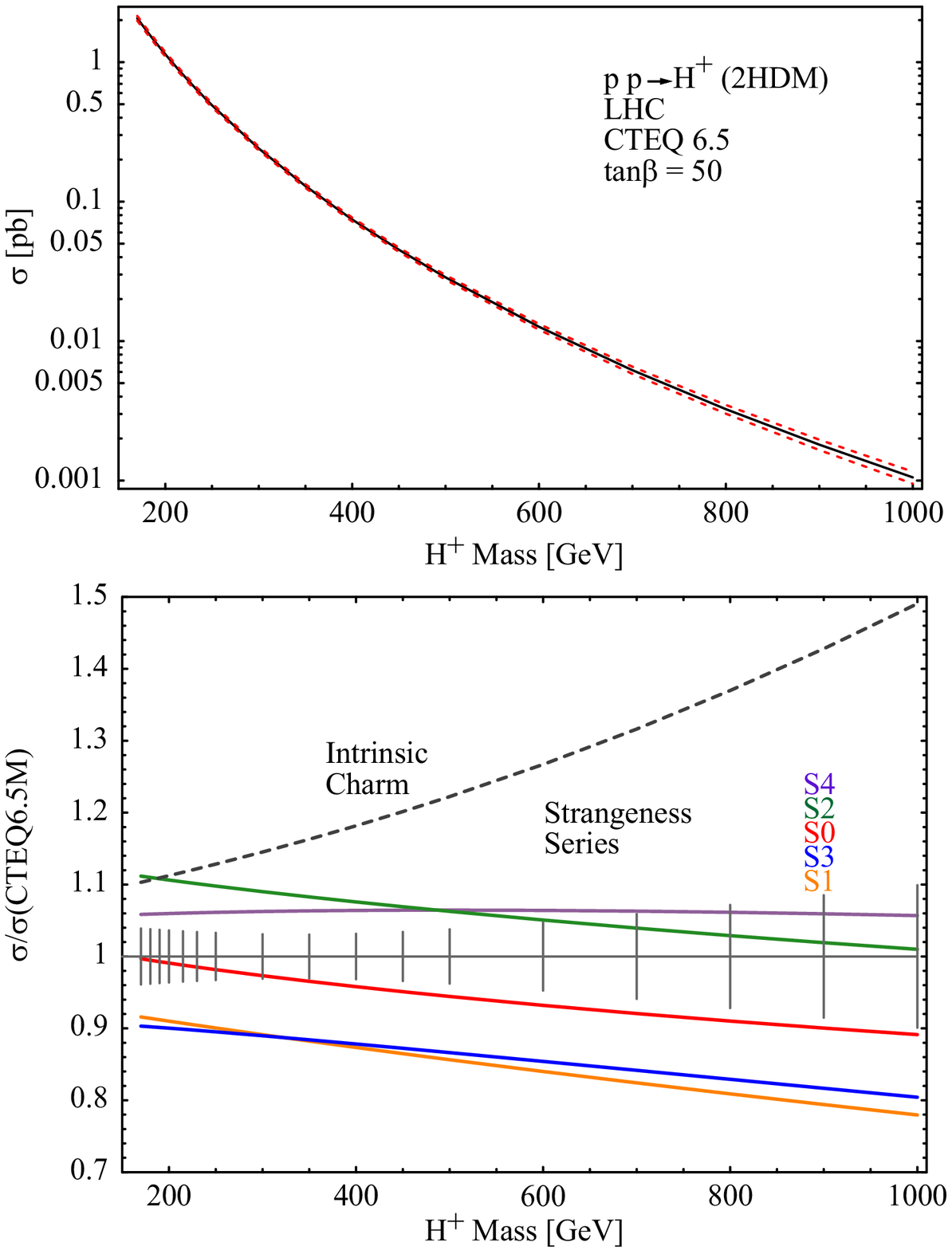}
 }
\centerline{\hspace{2em}(a)\hspace{0.5\textwidth}(b)}
\caption{
The total cross section for charged Higgs boson production in
 2HDM (top panel) and its fractional uncertainty (bottom panel) at (a)
the Tevatron Run-2 and (b) the LHC.
}%
\label{fig:Higgs}
\end{figure}
}
\newcommand{\figNutev}
{
\begin{figure}[h]
\hfill
 \resizebox*{0.5\textwidth}{!}{
 \includegraphics{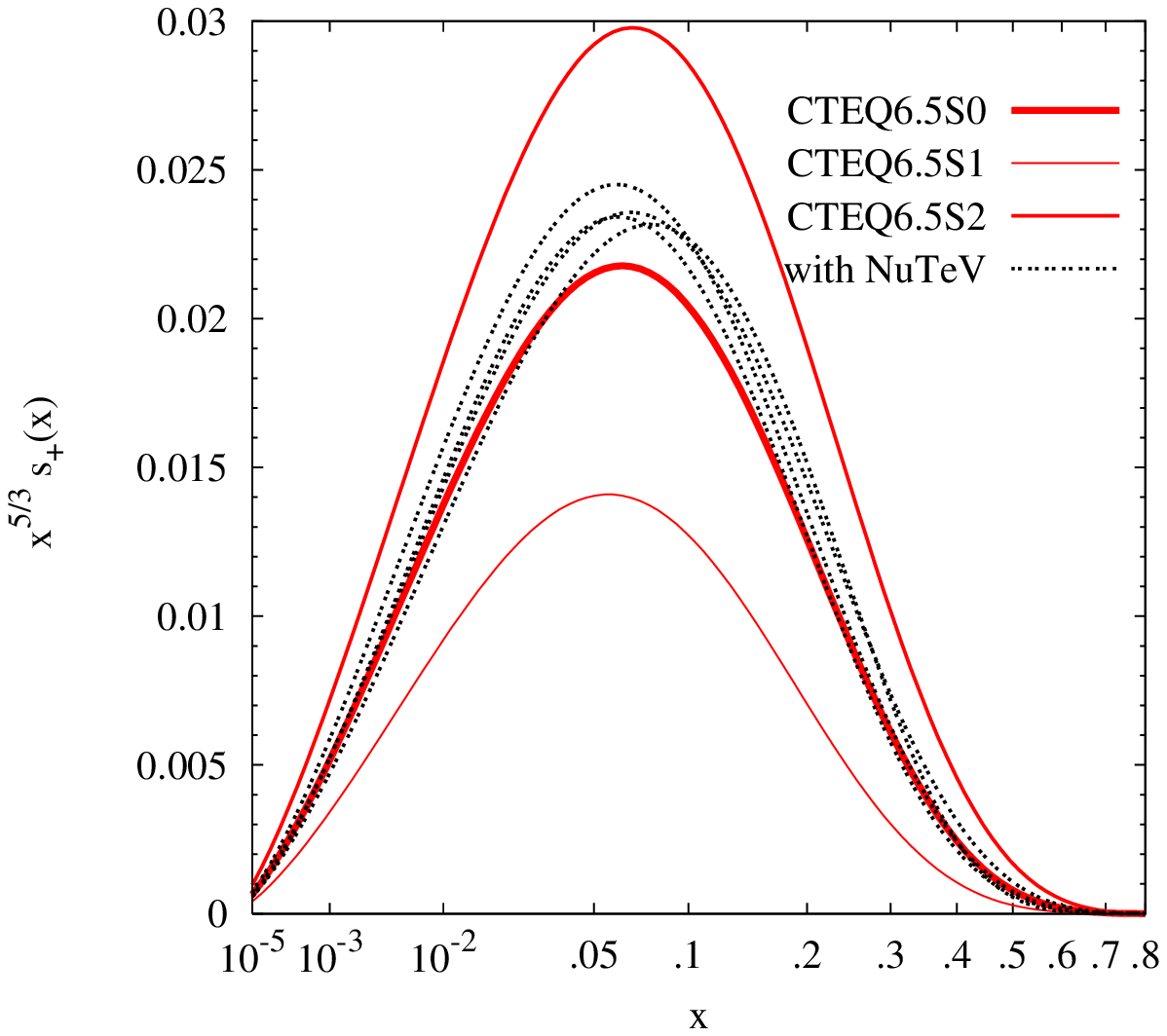}
 }
\hfill\rule{0em}{0em}
\caption{Typical symmetric strangeness sea $s_+(x)$ obtained using the NuTeV total inclusive
data, represented by dotted lines, compared to CTEQ6.5S0,1,2 PDFs (solid curves).
}%
\label{fig:nutev}
\end{figure}
}
\newcommand{\tblA}
{
\begin{table}[h]
\centering
\begin{tabular}{|c|c|c|c|}
\hline
change of & \multicolumn{3}{c|}{\# new parameters\rule{0em}{2.6ex}} \\
\cline{2-4}
goodness-of-fit & \makebox[2em]{1} & \makebox[2em]{2} & \makebox[2em]{3} \\
\hline \rule{0em}{2.8ex} $-\Delta \chi _{\mathrm{global}}^{2}$ (3542 pts.) &
65 & 68 & 69 \\ \hline \rule{0em}{2.8ex} $-\Delta \chi _{\mu ^{+}\mu
^{-}}^{2}$ (149 pts.) & 46 & 48 & 50 \\ \hline
\end{tabular}
\caption{Reduction in $\chi^{2}$'s with respect to the
reference fit CTEQ6.5M due to the introduction of new $s_+(x)$ strangeness
degrees of freedom.\label{tbl:s+} }
\end{table}
}
\newcommand{\tblB}
{
\begin{table}[h]
\centering
\begin{tabular}{|c|c|c|c|}
\hline
change of & \multicolumn{3}{c|}{\# of parameters\rule{0em}{2.6ex}} \\
\cline{2-4}
goodness-of-fit & \makebox[2.5em]{3} & \makebox[2.5em]{4} & \makebox[2em]{5} \\
\hline \rule{0em}{2.8ex} $-\Delta \chi _{\mathrm{global}}^{2}$ (3542 pts.) &
15 & 16 & 16 \\
\hline \rule{0em}{2.8ex} $-\Delta \chi _{\mu ^{+}\mu^{-}}^{2}$ (149 pts.) &
15 & 15 & 16 \\ \hline
\end{tabular}
\caption{Reduction in $\chi^{2}$ with respect to the reference
fit CTEQ6.5S0 due to the introduction of strangeness asymmetry parameters.
\label{tbl:s-}
}
\end{table}
}

\newcommand{\DATE}  {\today}

\newcommand{\PPrtNo}
{ ANL-HEP-PR-07-14, MSU-HEP-07012 }
\newcommand{\TITLE}
{The Strange Parton Distribution of the Nucleon: \\
 Global Analysis and Applications }

\newcommand{\AUTHORS}
{H.L.~Lai\footnote{Email: hllai@tmue.edu.tw}$^{a,b,c}$, P.~Nadolsky$^d$,
J.~Pumplin$^a$, D.~Stump$^a$, \\
W.K.~Tung$^{a,b}$, C.-P.~Yuan$^a$}
\newcommand{\INST}
{
$^a$ Michigan State University, E. Lansing, MI, USA \\
$^b$ University of Washington, Seattle, WA, USA \\
$^c$ Taipei Municipal University of Education, Taipei, Taiwan \\
$^d$ Argonne National Laboratory, Argonne, IL, USA}
\newcommand{\ABSTRACT}
{The strangeness degrees of freedom in the parton structure of the nucleon
are explored in the global analysis framework, using the new CTEQ6.5
implementation of the general mass perturbative QCD formalism of Collins. We
systematically determine the constraining power of available hard scattering
experimental data on the magnitude and shape of the strange quark and
anti-quark parton distributions. We find that current data favor a distinct
shape of the strange sea compared to the isoscalar non-strange sea.  A new
reference parton distribution set, CTEQ6.5S0, and representative sets
spanning the allowed ranges of magnitude and shape of the strange
distributions, are presented. Some applications to physical processes of
current interest in hadron collider phenomenology are discussed.}

\begin{document}


\begin{titlepage}

\noindent
\begin{tabular}{l} \DATE
\end{tabular}
\hfill
\begin{tabular}{l}
\PPrtNo
\end{tabular}

\vspace{1cm}

\begin{center}
\renewcommand{\thefootnote}{\fnsymbol{footnote}}
{
\LARGE \TITLE
}

\vspace{1.25cm}
{\large  \AUTHORS}

\vspace{1.25cm}

\INST
\end{center}

\vfill

\ABSTRACT                 

\vfill

\newpage
\end{titlepage}

\setcounter{footnote}{0}
\tableofcontents
\newpage



\section{Introduction\label{sec:intro}}

Although the global analysis of parton distribution functions (PDFs) has been
in progress for over two decades, certain components of the parton structure
of the nucleon are still poorly determined \cite{wkt04HCP,Thorne:2006wq}.
Foremost among these, perhaps surprisingly, is the strangeness sector. The
lack of knowledge on the strange and anti-strange parton distributions,
$s(x)$ and $\bar{s}(x)$, is reflected in the common practice in current
global QCD analyses of PDFs of adopting the simplifying ansatz $s = \bar{s} =
r \, (\bar{u}+\bar{d})/2$ (where $\bar{u}$ and $\bar{d}$ are the up and down
sea anti-quarks) at some low factorization scale. Even the proportionality
constant $r$ is only very loosely constrained by data.\footnote{This
uncertainty is not included in uncertainty measures such as the eigenvector
sets of CTEQ6, 6.1, 6.5, because the proportionality constant was held fixed
at a particular medium value for all those fits.}

Progress in incorporating heavy quark mass effects into the general
PQCD formalism, combined with recent precision cross section data
from HERA and fixed-target experiments, provides us with an
opportunity to re-assess the situation. This is important
theoretically, because the parton structure of the nucleon,
including strangeness, is a fundamental aspect of QCD dynamics at
the confinement scale.  It is also important phenomenologically,
because $s(x)$ and $\bar{s}(x)$ are significant for quantitative
calculations of certain key short-distance processes at hadron
colliders.

In a recent paper \cite{Cteq65m}, hereafter referred to as CTEQ6.5M,
we investigated how the up/down-quark and the gluon distributions,
and their uncertainties, are affected by the new theoretical and
experimental input, while keeping the conventional $s=\bar{s}
\propto ( \bar{u}+\bar{d})$ ansatz.  We found that the new
developments give rise to notable shifts in the PDFs which have a
significant impact on hadron collider phenomenology. In this work,
we extend the analysis to focus on the strangeness sector.
Specifically, we explore to what extent we can now quantify $s(x)$
and $\bar{s}(x)$ without restricting their shapes; and then consider
some implications of the improved PDFs on the phenomenology of
Standard Model (SM) and Beyond Standard Model (BSM) physics at the
Tevatron and the LHC.

In Section~\ref{sec:global}, we discuss issues relevant to
performing global QCD analysis with a focus on the strangeness
sector. In Section~\ref{sec:splus}, we study the symmetric strange
sea combination $s(x)+\bar{s}(x)$. We determine the number of new
strangeness parameters that can be constrained by current data,
present an improved PDF set CTEQ6.5S0 that best represents the
global data, and investigate the ranges of the magnitude and shape
of the symmetric strange sea allowed by the available constraints.
Several PDF sets that exemplify these ranges, CTEQ6.5Si
($i=1,\dots,4$), are given. In Section~\ref{sec:sminus}, we examine
the strangeness asymmetry function $s(x)-\bar{s}(x)$. We discuss
constraints that can be placed on this function by current global
analysis, and compare the results with existing ones.%
\footnote{In a previous study of the strangeness asymmetry as a
possible explanation of the NuTeV anomaly \cite{StrAsym03}, charm
quark mass effects were only applied to the calculation of the CC
dimuon production process. In the current study, the general mass
formalism is applied to all DIS processes in an unified way.} %
To illustrate the use of these new results, we consider in
Section~\ref{sec:apps} implications of the new PDFs on: (i)  the SM
process $p\,p\mkern-11mu \strut^{\scriptscriptstyle{(-)}} \to g \,+
\,\bar{s} \to W^{+} \,+\, c$; and the BSM process $p\,p\mkern-11mu
\strut^{\scriptscriptstyle{(-)}} \to \bar{s}\,+\,c\to H^{+}$. In
Section~\ref{sec:summary}, we summarize our results.



\section{Global Analysis with Focus on the Strange Sector\label{sec:global}}

The staple input to standard global QCD analyses of PDFs are the total
inclusive cross sections in deep inelastic scattering (DIS), Drell-Yan
processes (DY), and inclusive jet production.  These processes are largely
insensitive to the strange content of the nucleon, since the contributions
from the non-stange sea-quarks and gluon partons are larger and are of
similar shape to the strange sea distribution. In principle, at leading order
(LO) in QCD, the neutral current (NC) and charged current (CC) DIS structure
functions depend on somewhat different combinations of strange and
non-strange partons. Therefore certain specific combinations of these
structure functions can be sensitive to $s(x)$ and $\bar{s}(x)$. But in
practice, due to the large uncertainties inherent in combining data from
different kinds of experiments (CC vs.~NC), the constraints on strange and
anti-strange PDFs provided by total inclusive data are known to be very weak,
as we will also see quantitatively in this paper.

The semi-inclusive CC process
$\nu /\bar{\nu} \, + \, N \to \mu^{-/+} \, + \, H_{c} \, + \,X$
(where $N$ denotes a nucleon or nucleus and $H_{c}$ a charm
meson or baryon) is sensitive to strange distributions through the LO
partonic process $W^{+} \, + \, s \to c$, where $c$ is the charm quark.
The experimental signature of this process is dimuon production in neutrino
(anti-neutrino) scattering, since the semi-leptonic decay of the charm meson
gives rise to a second muon in the final state \cite{Goncharov:2001qe}. Our
study is partly motivated by the recent availability of the final analysis
of the high statistics NuTeV measurement of this process \cite{Mason:2006qa}.

The task of reliably constraining $s(x)$ and $\bar{s}(x)$ must be performed,
however, within the context of a comprehensive global QCD analysis. This is
because, beyond the LO in perturbative theory, QCD dynamics strongly couples
the strange degrees of freedom to the gluon and the other quark flavors. No
parton flavor can be determined in isolation: all available high precision
inclusive data that constrain the light degrees of freedom are needed in the
analysis. Furthermore, the charm final state in the dimuon production process
requires that heavy quark mass effects be included properly. Thus, a
consistent formalism incorporating mass effects in both QCD factorization
(dynamics) and in the phase space treatment (kinematics) must be applied to
all the relevant DIS processes in the global analysis. The global analysis
framework of CTEQ6.5 \cite{Cteq65m} contains these features, and thus
provides a suitable basis for performing this study.

It is convenient to work with symmetric and anti-symmetric combinations
of $s$ and $\bar{s}$:
\begin{equation}
s_{^{\pm }}(x,\mu )\,\equiv \, s(x,\mu )\, \pm \, \bar{s}(x,\mu ) \; .
\label{spmdef}
\end{equation}%
These have opposite CP properties and evolve differently under QCD evolution.
In Ref.\,\cite{Cteq65m}, the conventional ansatz
\begin{eqnarray}
s_{+}(x) &=&r[\bar{u}(x)+\bar{d}(x)]  \label{ansatz1} \\
s_{-}(x) &=&0  \label{ansatz2}
\end{eqnarray}%
at the initial scale $\mu = Q_{0} = 1.3 \, \mathrm{GeV}$ was made. In this
common approach, the only parameter associated with the strangeness degree of
freedom is the constant ratio $r$ of the strange to non-strange sea
distributions at $Q_0$.

In the current work, we extend the analysis to include additional \emph{%
independent parameters characterizing the shape of the strangeness
distributions}, and study their effect on the global fit. The number of new
parameters and the associated functional forms will be specified in
subsequent sections devoted to $s_+(x)$ and $s_-(x)$.

The experimental input to this study is essentially the same as in the
CTEQ6.5M analysis \cite{Cteq65m}. It consists of the full range of HERA I NC
and CC total cross section and semi-inclusive charm and bottom production
data, along with standard fixed-target NC and CC DIS and DY experiments, and
the Tevatron measurements of inclusive jet production and the charge
asymmetry of leptons from $W$ production.  Correlated systematic errors,
whenever available, are incorporated in the analysis.

Since the neutrino dimuon cross sections are of special importance in
exploring the strangeness degrees of freedom, we include both the CCFR and
the NuTeV data sets \cite{Goncharov:2001qe} for this analysis.%
\footnote{The same nuclear target corrections are made to the total
CC inclusive and the dimuon production data. We use the empirical
correction factors determined in NC lepton-nucleon and
lepton-nucleus scattering \cite{nmcNc}. See the related discussions in
the appendix on other possible choices of nuclear
corrections concerning the NuTeV total cross section data.} %
The measured ``forward dimuon production'' cross sections from
\cite{Goncharov:2001qe} cannot be directly compared to our
theoretical calculation of inclusive charm production cross section,
because they also depend on the fragmentation of the charm quark
into charmed particles, and the decay of those particles. We rely on
the results of the recent comprehensive analysis of Mason
\cite{Mason:2006qa} to make the connection between the two.%
\footnote{We thank David Mason for providing the results of his analysis, and
for detailed discussions concerning the proper use of these results.}

In the global analysis of PDFs, the parton degrees of freedom must be matched
to the constraining power of the available input experimental data within the
adopted theoretical framework, in order for the results to be meaningful. The
first question in studying the strangeness sector of the nucleon parton
structure must therefore be: are current theory and experiment able to
discern independent non-perturbative strange degrees of freedom; and, if so,
how many such degrees of freedom are needed?

We shall try to answer this question by comparing global fits of the
conventional type (with no independent strange parton degrees of freedom
except the overall normalization) to new fits obtained with various numbers
of new strangeness shape parameters. The reference PDF set, with the
conventional ansatz (\ref{ansatz1})--(\ref{ansatz2}) imposed at $\mu
=Q_{0}=1.3\,\mathrm{GeV}$, is essentially CTEQ6.5M \cite{Cteq65m}.



\section{The Symmetric Strange Sea $\mathbf{s_{+}(x)}$\label{sec:splus}}

To extend the analysis to the strangeness sector of parton parameter space,
we first examine the symmetric component of the strange sea, $s_{+}(x)$. For
the non-perturbative input distribution, we adopt the standard initial form
\begin{equation}
s_{+}(x,Q_{0})=A_{0}\,x^{A_{1}}\,(1-x)^{A_{2}}\,P_{+}(x) \ \ ,
\label{splusdef}
\end{equation}%
where $P_{+}(x)$ is a smooth positive definite function on the interval $%
(0,1)$. The function $P_{+}(x)$ depends on additional parameters $%
A_{3},\dots $ as needed. The normalization constant $A_{0}$ which controls
the \emph{overall magnitude} of the strange distribution is related to the
strange/non-strange ratio parameter $r$ in Eq.~(\ref{ansatz1}), and hence it
will not be counted as a new parameter. The parameters $A_1,A_2,\dots$ define
the shape of the non-perturbative strange distribution $s_{+}(x,Q_{0})$.
Because the experimental constraints on the new parameters are not tight, we
generally retain the relation $A_{1}^{s_{+}}=A_{1}^{\bar{u}+\bar{d}}$,
suggested by common Regge considerations for small-$x$ behavior of PDFs, to
reduce arbitrariness.%
\footnote{%
Relaxing this constraint does not lead to any meaningful improvement in the
global fit.} This leaves $A_{2},A_{3},\dots $ as the new strangeness
parameters to be investigated in the rest of this section.

\subsection{The Number of New Shape Parameters \label{sec:numberofpar}}

We have examined the quality of global fits obtained with a variety of
functional forms for $P_{+}(x)$ that involve new shape parameters, compared
to CTEQ6.5M which is taken as a reference fit with $s_{+}
\propto (\bar{d} + \bar{u})$. The main results are summarized in Table \ref%
{tbl:s+}, which shows the reduction in $\chi ^{2}$ with respect to CTEQ6.5M
(hence the improvement in the goodness-of-fit) for the full data set $\Delta
\chi _{\mathrm{global}}^{2}$ and for the dimuon data sets $\Delta \chi _{\mu
^{+}\mu^{-}}^{2}$, when $1$, $2$, or $3$ new strangeness parameters are added
to the global fit. The $1$ new parameter case---with $A_2$ only---corresponds
to $P_+=1$ in Eq.\,(\ref{splusdef}); the other cases involve various
choices of non-trivial functions  $P_+(x)$.\footnote{%
A representative functional form is $P_{+}(x)=e^{A_{3}\sqrt{x}
+A_{4}x+A_{5}x^{2}}$, with one or more of the $A_{i}$ set to zero.
The exponential form ensures positivity of the parton distribution.}

\tblA%
We see that whereas, at first sight, the significance of $\Delta \chi _{%
\mathrm{global}}^{2}\sim 67$ for the full global data set of $3542$
points may be arguable, it is striking that the bulk of
this reduction is concentrated in the most physically relevant $\nu $ and $%
\bar{\nu}$ dimuon data sets. The improvement in the goodness-of-fit for the
dimuon data sets---$\Delta \chi _{\mu ^{+}\mu ^{-}}^{2}\sim 48$---is quite
significant, because, for 149 data points, the normal 90\% confidence level
criterion corresponds to a $\Delta \chi^2$ of 22. Our general analysis
procedure, as described in more detail in \cite{Cteq65m}, requires acceptable
fits to be within the 90\% confidence levels of all contributing data sets.
In the current study, the deciding
factor is therefore the goodness-of-fit of the dimuon data sets.%
\footnote{The slight reduction of $\Delta \chi^{2} \sim 20$, spread evenly
among the remaining $\sim 3400$ data points, is consistent with the
preference of the new degrees of freedom, but it is not significant by
itself.} %
This was already anticipated in Sec.\,\ref{sec:global}.
Fig.\thinspace \ref{fig:nutevS} gives a graphical illustration of
the improvement on the fit to the more accurate neutrino dimuon data
sets using a new PDF set with (minimal) independent strangeness
sector (cf.\ next subsection) compared to that using CTEQ6.5M.
\fignutevS

We conclude from these results that \emph{the current global analysis
strongly favors a different shape for the strange sea} $s_{+}(x)$, \emph{%
compared to the (isoscalar) non-strange sea}
$\bar{u}(x)+\bar{d}(x)$. From the physics point of view, this is
only natural, since the initial parton distributions reflect low
energy non-perturbative physics. QCD dynamics at long distances does
differentiate the strange and non-strange sectors, as seen in hadron
spectroscopy. The conventional ansatz of \emph{the same shape} but
\emph{different sizes} for the initial parton distributions has been
only a convenient working ansatz in the absence of sufficient
experimental constraints on the difference. The results presented
above show that recent improvements in theory and experiment now
allow us to discern the difference.

However, Table \ref{tbl:s+} also shows the limitation of current
experimental constraints: \emph{there is no significant improvement in the
goodness-of-fit when the number of new degrees of freedom is increased
beyond 1}. This means our global analysis cannot distinguish the simple shape%
\begin{equation}
s_{+}(x,Q_{0})=A_{0}\,x^{A_{1}}(1-x)^{A_{2}}  \label{splus1}
\end{equation}%
from more complex forms that invoke additional parameters
($P_{+}(x;\,A_{3},A_{4},A_{5})\neq 1$ in Eq.\,(\ref{splusdef})).
Because of this, Eq.\,(\ref{splus1}) will suffice as a starting
point for exploring the strangeness sector of parton parameter
space.

\subsection{New Central PDF set, CTEQ6.5S0 \label{sec:c65s0}}

With the independent strangeness sector represented by the input function (%
\ref{splus1}), we need to establish a new reference PDF set that
provides the best fit to the current global hard scattering data.
This set will be referred to as CTEQ6.5S0 in subsequent discussions.
It represents an improvement of the CTEQ6.5M PDF set of
\cite{Cteq65m} as discussed in the
last subsection. Except for the differentiation between $s_{+}(x)$ and $\bar{u%
}(x)+\bar{d}(x)$ shapes, these two PDF sets are very similar to each other.

In Fig.\thinspace \ref{fig:6m6sComp} we compare the sea
distributions from the two PDF sets. The $x$-axis is scaled as
$x^{1/3}$ so that the large-$x$ and small-$x$ regions are both
clearly displayed. The $y$-axis is scaled so that the area under
each curve is proportional to the momentum fraction carried by that
PDF. We see that the isoscalar non-strange sea
$\bar{u}(x)+\bar{d}(x)$ is almost unchanged, while the new CTEQ6.5S0
symmetric strange sea $s_{+}(x)$ has become somewhat smaller and
softer compared to CTEQ6.5M.\figmsComp

\subsection{Constraints on the Magnitude of $s_{+}$\label{sec:magnitude}}

The \textquotedblleft magnitude\textquotedblright\ of a parton flavor $f$ ($%
= g,u,\bar{u},d,\bar{d},s,\bar{s},...$) inside the nucleon is naturally
represented by the momentum fraction it carries:\footnote{%
The parton number integral $\int_{0}^{1}f(x)\,dx$ does not converge
for most parametrized forms of the gluon and the sea quarks, so it
\emph{does not} make an appropriate measure of the overall size of a
flavor component.}
\begin{equation}
\langle x\rangle _{f}\equiv \int_{0}^{1}x\,f(x)\,dx\ \ .  \label{momfrac}
\end{equation}%
By this measure, the value of $\langle x\rangle _{s_{+}}$ for the reference
CTEQ6.5S0 set is $0.027$, while for CTEQ6.5M it was $0.032$.

To study the uncertainty in the magnitude of $s_{+}$, we performed a series
of global fits using the Lagrange Multiplier method of \cite{MsuMV,MsuLgr} to
map out the allowed range of $\langle x\rangle _{s_{+}}$. Keeping in mind the
results of Sec.\thinspace \ref{sec:numberofpar}, as the value of $\langle
x\rangle _{s_{+}}$ is varied around the central value $0.027$, we keep track
of the variation in overall $\chi _{\mathrm{global}}^{2}$ as well as the
variation in $\chi _{\mu ^{-}\mu ^{+}}^{2}$. The results are presented in
Fig.\thinspace \ref{fig:StrChiKappa}.%
\figChi For convenience, the comparison is made on $\chi ^{2}$ per data
point, with the number of data points being given in Table \ref{tbl:s+}. We
see that, as expected, the $\nu $ and $\bar{\nu}$ dimuon data sets are quite
sensitive to the magnitude of the strangeness PDFs, whereas the rest of the
global data sets are basically insensitive to it.

Thus, the experimental constraints on $\langle x\rangle _{s_{+}}$
are provided essentially by the $\nu $ and $\bar{\nu}$ dimuon data
sets. \ Following Refs.\thinspace \cite{Cteq6,Cteq65m}, we determine
the uncertainty range of $\langle x\rangle _{s_{+}}$ by the 90\%
confidence criteria on the dimuon production data sets. This range
is $0.018\,<\langle x\rangle _{s_{+}}<\,0.040\,$. The two sets of
PDFs that represent the best fits corresponding to the lower (upper)
bound value of $\langle x\rangle _{s_{+}}$ will be referred to as
CTEQ6.5S1 (CTEQ6.5S2). The variation of $\langle x\rangle _{s_{+}}$
is strongly correlated with the normalization of the $\nu $ and
$\bar{\nu}$ dimuon production data sets compared to the
theoretically calculated inclusive charm production cross section.
There are various sources of uncertainty on this overall factor:
experimental (global and energy-dependent) normalization of the
total cross sections ($\sim 2-5\%$), fragmentation function of charm
quark to charmed hadrons, branching ratio of charmed hadron decay to
muon ($\sim 10\%$), \dots, etc. These are taken into account
according to our standard uncertainty analysis. The limits of
$\langle x\rangle _{s_{+}}$ obtained above correspond to $\pm 20\%$
ovreall variation of the normalization factor, as determined by this
analysis procedure. As the magnitude of $\langle x\rangle _{s_{+}}$
varies, the shape of $s_{+}(x)$ also adjusts to best fit the global
data. A plot of $s_{+}(x)$ for these PDFs will be shown in the next
section.

The \emph{relative strength} of the strange partons is conveniently
characterized by the ratio of $\langle x\rangle _{s+\bar{s}}$ to the average
of strange and non-strange sea $\kappa \equiv 3\langle x\rangle _{s+\bar{s}%
}/(2\langle x\rangle _{\bar{u}+\bar{d}}+\langle x\rangle _{s+\bar{s}})$
(Note, $u_{\mathrm{sea}}=\bar{u}$, $d_{\mathrm{sea}}=\bar{d}$.) The values
of $\kappa $ for CTEQ6.5Si, $i=1,0,\,2$, are $0.35,\,0.54,\,0.75,$
respectively. The relative size of the strange and non-strange distributions
can also be expressed in terms of the ratio of strange to non-strange sea, $%
r=\langle x\rangle _{s+\bar{s}}/\langle x\rangle _{\bar{u}+\bar{d}}=2\kappa
/(3-\kappa )$. This is a generalization of the ratio parameter introduced in
Eq.\thinspace (\ref{ansatz1}) for the special case of the same shape for
strange and non-strange seas. The PDF sets CTEQ6.5Si, $i=1,\,0,\,2$,
correspond to $r=0.27,\,0.44,0.67$, respectively; while $r = 0.50$ for
CTEQ6.5M.

\subsection{Constraints on the Shape of $s_{+}(x)\label{sec:shape}$}

It is also of interest to ask what is the uncertainty in the \emph{shape} of
the input $s_{+}(x)$. To study this, we need to expand the input
functional form beyond the minimal one given in Eq.\thinspace (\ref{splus1}).
As shown in Table 1, fits obtained with $P_{+}(x)$ in Eq.\,(\ref{splusdef})
that contain additional parameters will have comparable
goodness-of-fit to current data compared with those described above (with $%
P_{+}(x)=1$). The variation of the $s_{+}(x)$ shape among fits obtained with
expanded parametrizations should therefore give a reasonable measure of the
range allowed by existing data.

To carry out this study, we explored a variety of choices for $%
P_{+}(x;A_{3},A_{4})$ and examined the shape of $s_{+}(x)$ for candidate
fits within the 90\% confidence level constraints of the dimuon data sets.
(The goodness-of-fit to the majority inclusive data sets was again
insensitive to these variations, staying consistently close to the optimal
level.) It is difficult to uniquely characterize the ``shape''
of a function such as $s_{+}(x)$ when the experimental constraints
are relatively weak. We have chosen two representative alternative PDF sets
among those explored just to illustrate the range allowed by the 90\%
confidence level criterion. These will be referred to as CTEQ6.5S3 and
CTEQ6.5S4.

Figure \ref{fig:splus5} shows the strangeness distribution $s_{+}(x)$ at the
initial scale $\mu =1.3\,\mathrm{GeV}$ for the five PDF sets, CTEQ6.5Si, $%
i=0,...,4$. The axes are scaled the same way as in Fig.\thinspace \ref%
{fig:6m6sComp}. The area under each curve corresponds to the momentum
fraction carried by $s_{+}$, hence directly illustrates its magnitude. The
solid (red) curve shows the reference PDF set CTEQ6.5S0. The other curves
illustrate the range of variation of the magnitude and shape of $s_{+}(x)$
allowed by current data. \figsplus



\section{The Strangeness Asymmetry $\mathbf{s_{-}(x)}$\label{sec:sminus}}

To study the anti-symmetric combination $s_{-}(x)$, the \emph{strangeness
asymmetry function}, we must bear in mind that \emph{(i)} if we want to keep
both $s$ and $\bar{s}$ parton distributions positive definite, it is
necessary to maintain the condition $|s_{-}|\leq |s_{+}|$; and \emph{(ii)} in
order to satisfy the strangeness quark number sum rule, we must have
$\int_0^1 s_{-}(x)dx=0$. The latter condition implies that $s_{-}(x)$ changes
sign at least once in the interval $0<x<1$. Current data do not have enough
discriminating power to establish multiple oscillatory behavior of $s_{-}(x)$%
, so we shall restrict the parametrization of $s_{-}(x)$ to the case of a
single crossing of the $x$-axis.

A convenient smooth parametrization that has the required features is
\begin{equation}
s_{-}(x,Q_{0})=s_{+}(x,Q_{0})\frac{2}{\pi}\tan^{-1} [\,c\,x^{a}(1-\frac{x}{b}%
)\,e^{dx+ex^{2}}]  \label{sminus}
\end{equation}%
where $c$ controls the sign and the overall magnitude of $s_{-}(x)$, $a$
characterizes the difference in small-$x$ behavior between $s_{-}(x)$ and
$s_{+}(x)$ (expected on physical grounds), $b$ represents the $x$-value at
which $s_{-}(x)$ changes sign, and the factor $e^{dx+ex^{2}}$ may be included
if needed to provide more flexibility in the shape of the function. We see
that, in order to have a non-trivial strangeness asymmetry, a minimum of 3
parameters---{$a,b,c$} in Eq.\,(\ref{sminus})---are required to characterize
the non-perturbative input function.

The exploration of the strangeness asymmetry sector can be pursued by the
same procedure as for $s_{+}(x)$. The experimental constraints are again
expected to come mostly from the $\nu $ and $\bar{\nu}$ dimuon production
data sets. \tblB The results of an extensive study are summarized in Table %
\ref{tbl:s-}. This time we use CTEQ6.5S0 as the reference fit, and examine
the improvement in goodness-of-fit due to the introduction of $3$ (the
minimum) to $5$ parameters to characterize strangeness asymmetry degrees of
freedom. The $3$-parameter case refers to the set {$a,b,c$}; and the other
cases add the parameters {$d,e$} in order. The numbers given in this table
are not unique since many equivalent examples have been studied; they are
representative of the general pattern.

\paragraph{Significance of non-zero Strangeness Asymmetry:}

Compared to the case of $s_{+}$ (Table \ref{tbl:s+}), we see that
the reductions in $\chi _{\mathrm{global}}^{2}$ (with respect to
CTEQ6.5S0 which has a symmetric strangeness sea) are insignificant
(for the total number of points), and all the reductions come from
the dimuon data sets. Furthermore, the numbers for $\Delta \chi
_{\mu ^{+}\mu ^{-}}^{2}$ are a factor of 3 smaller than those
appearing in Table \ref{tbl:s+}; and are below the 90\% C.L. figure
of 22. Thus, we consider the improvement on the goodness-of-fit over
the no strangeness asymmetry case to be marginal. This does not
mean, however, that experimental evidence is against the existence
of strangeness asymmetry! To the contrary, the latter can be sizable
since the constraints are shown to be weak. Thus, in the next
subsection, we shall investigate the allowed range of this
asymmetry, assuming it is non-zero.

One feature of these fits is worth noting: the fitted value for the $\,a\,$
parameter of Eq.\thinspace (\ref{sminus}), if let free, is generally in the
range of the theoretical expectation $\sim 0.5$---the difference of the two
Regge intercepts of the CP even and odd trajectories. In view of this fact,
in further studies described below, we normally set $a=0.5$ and let the other
parameters vary, in order to render the very loosely constrainted fits more
stable.

The magnitude of strangeness asymmetry can again be characterized by
the first moment of $s_{-}$: $\langle x\rangle
_{s_-}=\int_{0}^{1}x\,s_{-}(x,Q_{0})\,dx$. The best fit to global
data, using the minimal parametrization ($d\!=\!e\!=\!0$ in
Eq.\thinspace (\ref{sminus})), corresponds to $\langle x\rangle
_{s_-}=0.0018$.  This value is completely consistent with those of
the previous global analysis \cite{StrAsym03} and with the recent
final experimental analysis of the NuTeV dimuon data alone
\cite{Mason:2006qa}.

\paragraph{Range of Allowed Strangeness Asymmetry:}

The uncertainties of the magnitude and shape of $s_{-}(x)$ can again be
studied using the Lagrange Multiplier method. Applying the 90\% confidence
criterion, we estimate the range of the magnitude to be $-0.001<\langle
x\rangle _{s_-}<0.005$. This range again coincides with that found in \cite%
{StrAsym03}.

We note that $s_{-}(x)$ is particularly sensitive to the difference between
the $\nu $ and $\bar{\nu}$ cross sections. In our analysis, we treat the
overall normalization of the NuTeV $\nu $ and $\bar{\nu}$ dimuon data sets as
a fitting parameter (as already mentioned in Sec.\thinspace \ref{sec:splus}),
but keep the relative normalization between these data sets fixed. If this
relative
normalization were allowed to float, both the magnitude and the shape of $%
s_{-}(x)$ would change significantly (along with a reduction in $\chi^2$'s
that is larger than those shown in Table \ref{tbl:s-}).

In view of the large range of uncertainty on $\langle x\rangle
_{s_-}$, which includes the possibility of zero asymmetry, we do not
think a strong statement can be made about the strangeness
asymmetry. However, the results of existing phenomenological studies
(Refs.\,\cite{StrAsym03,Mason:2006qa} as well as the present work)
and physical considerations (e.g.~the light-cone wavefunction models
\cite{models}) all suggest that $\langle x\rangle _{s_-}$ is most
likely positive. Figure~\ref{fig:sminus} illustrates typical shapes
of the asymmetry function $s_-(x)$ and the corresponding momentum
distribution $x\,s_-(x)$. The central curve (solid line) corresponds
to the best fit; the two extreme curves represent the two limiting
cases corresponding to the lower and upper bounds of $\langle
x\rangle _{s_-}$ described above; and the other two provide
alternative examples. These choices represent different possible
small-$x$ behaviors that span the entire physically allowed range,
not just that conforming to the Regge lore. \figSminus



\section{Physical Applications\label{sec:apps}}

The variations of $s(x)$ and $\bar{s}(x)$ found in the previous sections
will affect predictions for physical processes that are sensitive to the
strange parton distributions. In this section, we will describe two
examples: one within and one beyond the SM. A more thorough study will be
the subject of a separate paper \cite{Impact07}.

\subsection*{$W+c$ production and the strangeness distribution}

The inclusive cross section for the hadron collider process
$pp\mkern-11mu \strut^{\scriptscriptstyle{(-)}} \to W+c+X$
is sensitive to the strange parton
distributions through the LO partonic process $g\,\,{s}\to W\,{c}$
\cite{KellerGiele,wkt04HCP}, where $W$ stands for $W^{\pm }$; $c$, $s$ for the
corresponding quark or antiquark. Consider the rapidity distribution $%
d\sigma /dy$ of the $W^{+}$ boson with the constraint $q_{T}>20$ GeV on the
transverse momentum of the outgoing $W^{+}$ boson in the
LO approximation (order $%
\alpha _{s}$).\footnote{%
In the leading-order $g\,{\bar{s}}\to W^{+}{\bar{c}}$ partonic
process, the condition $q_{T}>20 \, \mathrm{GeV}$ implies the constraint
$p_{T}^{c}>20 \, \mathrm{GeV}$ on the transverse momentum of the
final-state charm quark because of
transverse momentum conservation.} The cross section can be calculated
neglecting quark masses in the Wilson coefficient, since these are very small
compared to the typical energy scale of the $W^{+}{\bar{c}}$ system. The
range of variation of $d\sigma /dy$ for Tevatron Run-2, using the
CTEQ6.5Si input PDFs ($i=0-4$), is shown in Fig.\thinspace %
\ref{fig:gs2wc}(a). The cross section in the central rapidity region varies by $%
\sim 30\%$ among the candidate PDFs. This exceeds the PDF uncertainty
estimate obtained using the CTEQ6.5 eigenvector sets (all of which assume
the ansatz (\thinspace \ref{ansatz1}) with a fixed $r=0.5$). \figGsWc This suggests that the
measurement of this cross section at the Tevatron will provide useful
constraints on the strange distribution.

To make a more realistic
feasibility study, it will be necessary to include higher-order QCD contributions,
detector acceptance corrections, and background estimates.
Here we shall only briefly examine how NLO QCD corrections can be expected to
affect the above
results. The calculation is carried out by adapting the $\mathcal{O}(\alpha
_{s}^{2})$ cross section formulas for high-$q_{T}$ $W$ boson production from
Ref.~\cite{Arnold:1988dp}. The most important higher-order contributions to
the PDF uncertainty estimate come from the 1-loop corrections to the
subprocess $g\,{\bar{s}}\to W^{+}\,{\bar{c}}$, and the tree-level $%
2\to 3$ process $g\,g\to s\,W^{+}\,{\bar{c}}$. (Our NLO
contributions are computed with the constraint $q_{T}>20 \, \mathrm{GeV}$
only and may
slightly overestimate the NLO rate if a lower cut is imposed on $p_{T}^{c}$
to experimentally identify the $c$-quark jet.) The results are well
represented by an overall multiplicative factor of about 1.3 applied to the
LO rapidity distributions shown in Fig.~\ref{fig:gs2wc}, for all PDF sets.
Hence, the NLO correction preserves sensitivity of the inclusive rapidity
distribution to the strange PDF.

Figure\,\ref{fig:gs2wc}(b) shows the results of a similar
calculation for inclusive $W^{+}\,{\bar{c}}$ production at the LHC. We see
that the range of variation of the cross section is more modest in this
case. This is because the relevant $x$-range for the LHC has a smaller
variation of the strangeness PDFs (see Fig.\,\ref{fig:splus5}).

\subsection*{Charged Higgs boson production, strangeness, and intrinsic charm%
}

The production of a charged Higgs boson $H^{+}$ via the partonic process $c+{%
\bar{s}}\to H^{+}$ provides an example of a BSM process that is sensitive to
the strange PDF. A charged Higgs field arises in many models where
electroweak symmetry breaking involves two or more doublets of Higgs bosons.
We shall consider specifically the Type-2 two Higgs doublet
model \cite{HiggsHunter}. The $c\,\bar{s}%
\to H^{+}$ amplitude is enhanced compared to other quark flavors by
relatively large $\overline{MS}$ quark
masses $\overline{m}_{s}$ and $\overline{m}_{c}$,
substantial magnitudes of antistrange and charm PDFs, and the nearly maximal CKM
matrix element $V_{cs}\approx 1$. To illustrate the range of variation in
predictions for this process at both the Tevatron and the LHC, we compute
the on-shell $H^{+}$ production cross section as a function of $M_{H^{+}}$.
The calculation includes the NLO QCD correction \cite{HeYuan}
and is evaluated for $\tan
\beta =50$, $\overline{m}_{s}(M_{Z})=0.084 \, \mathrm{GeV}$, $\overline{m}%
_{c}(M_{Z})=0.74 \, \mathrm{GeV}$, assuming 3-loop running of
$\overline{m}_{q}(\mu )$ in QCD.\footnote{%
The rates for a different value of $\tan \beta $ can be easily obtained by a
simple scaling factor $(\tan \beta /50)^{2}$.}

The upper frame in Fig.~\ref{fig:Higgs}(a) shows the CTEQ6.5
cross section for $H^{+}$ boson production in the Tevatron Run-2, including
the central CTEQ6.5M prediction (solid line), and the uncertainty band
obtained with the eigenvector sets (bounded by the dashed lines). The lower
frame shows the predictions for the strangeness PDF series CTEQ6.5Si, $i=0-4$%
, as ratios to the CTEQ6.5M cross section. The CTEQ6.5 uncertainty band is
shown as vertical lines. We observe that the strangeness distributions
CTEQ6.5Si generally predict smaller cross sections.  The decrease in the
prediction for the $H^{+}$ cross section can be as much as 20\% at small $%
M_{H^{+}}$ and 50\% at large $M_{H^{+}}$.

\figHiggs

The parton luminosity in the $c\,\bar{s}\to H^{+}$ process depends
also on the behavior of the charm parton distribution. The CTEQ6.5Si series
is generated under the conventional assumption that charm (anti)quarks are
only perturbatively generated. Charm distributions of a non-perturbative
origin---intrinsic charm (IC)---is both a theoretical possibility and
phenomenologically allowed, as discussed recently in \cite{Cteq65C}. The
existence of IC can significantly enhance the $H^{+}$ production cross
section at the Tevatron. The dashed curve in the ratio plot of Fig.\ \ref%
{fig:Higgs} represents the prediction obtained from a PDF set with a
conventional strange input, but an intrinsic charm whose shape is
given by the BHPS model \cite{BHPS,Cteq65C}. This dashed curve
reflects the largest magnitude of the intrinsic charm contribution
with $\langle x\rangle_{c+\bar c} =2.0\%$  allowed by the current
hard scattering data \cite{Cteq65C}. Thus the enhancement in $H^{+}$
production due to intrinsic charm may be as large as a factor of
2--3, with important implications for the Tevatron experiments. A
useful process to constrain the intrinsic charm contribution is the
associated $Z\,c$ production, which will be discussed in another
paper \cite{Impact07}.

Figure~\ref{fig:Higgs}(b) shows analogous plots for the
$c\,\bar{s}\to H^{+}$ cross sections at the LHC. In this case,
variations in the strangeness content may change the cross section in the
shown range of Higgs boson masses by $\pm 10-20\%$. The intrinsic charm
contributions may enhance the cross section by $10-50\%$. If $H^{+}$ is
observed, the measurement of its production rate would provide clues about
the underlying dynamics beyond the standard model. But for this test to be
successful, uncertainties in strange and charm distributions must be reduced
via independent measurements, such as dedicated studies of $W\,c$ and $Z\,c$
production.



\section{Summary\label{sec:summary}}

We have systematically investigated the strangeness sector of the parton
structure of the nucleon, using the latest theoretical tools and experimental
input, within the framework of precision global QCD analysis. We find strong
evidence that the strange quark distribution has a different shape compared
to that of the isoscalar non-strange sea quark distribution.

We studied the range of possible magnitudes and shapes of the symmetric
strange distribution function $s_{+}(x),$ using a 90\% confidence level
criterion. The range of allowed magnitude for the strange distribution, as
measured by the average fractional momentum $\langle x\rangle _{s_{+}}$, is
determined to be $0.018$--$0.040$. We present PDF sets that are
representative of this range.

We find that the strangeness asymmetry function $s_{-}(x)$ is only loosely
constrained. The range on the magnitude of this asymmetry, as measured by the
momentum integral is determined to be $-0.001 < \langle x\rangle _{s_{-}} <
+0.005$, which is consistent with results in the existing literature.

We also discuss two sample applications of the PDFs presented in
this study: the SM process $p\,p\mkern-11mu
\strut^{\scriptscriptstyle{(-)}} \to W^{+} + \,\bar{c}\, +X$ that
can help further constrain the strange distributions; and the BSM
process $p\,p\mkern-11mu \strut^{\scriptscriptstyle{(-)}} \to H^{+}
 +  X$, which depends sensitively on the strange and charm content
of the nucleon.

The new PDFs discussed in this paper, CTEQ6.5Si, $i=0-4$, will be made
available at the CTEQ web site (http://cteq.org/) and through the LHAPDF
system\\ (http://hepforge.cedar.ac.uk/lhapdf/).

\paragraph{Acknowledgment}

We thank Joey Huston for discussions and helpful comments.  This work was
supported in part by the U.S. National Science Foundation under awards
PHY-0354838 and PHY-0555545; US Department of Energy, Division of High Energy
Physics, Contract DE-AC02-06CH11357; and by the National Science Council of
Taiwan under grant NSC-95-2112-M-133-001.

\appendix
\section*{Appendix}
There are currently some unresolved problems among fixed-target total
inclusive neutrino and anti-neutrino scattering experiments on nuclear
targets by CDHSW \cite{CDHSW}, CCFR \cite{CCFR2,CCFR3}, NuTeV
\cite{NuTeVxsc}, and CHORUS \cite{ChorusSFxsc}, particularly in the large $x$
region. Comparisons of these experimental results can be found in
Refs.~\cite{ChorusSFxsc} and \cite{TzanovDis05}. A detailed study of these
problems in the global analysis context, including effects of nuclear target
corrections, has recently been performed in \cite{JeffEtal}. This open issue
does not affect our study of the strange sector, since the strange
distributions are insensitive to total inclusive cross sections, as
demonstrated in Sections \ref{sec:splus} and \ref{sec:sminus}.

Specifically, the results presented in the main body of this paper were
obtained by global analyses using fixed-target CC DIS data of CDHSW, CCFR,
and CHORUS, which are mutually consistent. Separate fits were also done using
the NuTeV \emph{total cross section} data in place of these.  Our findings
can be summarized as follows: (i) these fits uniformly have much worse
goodness-of-fit, to the point of beyond the 90\% C.L. for many NC DIS data
sets (BCDMS, H1, ZEUS); and (ii) if the resulting PDFs are taken at face
value, all our conclusions about the (not so well-constrained) strange parton
distributions remain valid, even as the better known $u,d,g$ distributions
are shifted from previous determinations. Fig.\thinspace \ref{fig:nutev}
shows four typical $s_{+}(x)$ curves obtained using the NuTeV data set with
different nuclear target corrections, compared to the CTEQ6.5S0,1,2 ones. We
see that they are well within the range determined with our default data
sets.

\figNutev


\end{document}